\documentclass[twocolumn,aps,prx,superscriptaddress]{revtex4}
\usepackage{graphicx}
\usepackage{color}
\usepackage{amsmath,amssymb}

\newcommand{\bra}[1]{\langle#1|}
\newcommand{\ket}[1]{|#1\rangle}

\begin{document}

%%%%%%%%%%%%%%%%%%%%%%%%%%%%%%%%%%%%%%%%%%%%%%%%%%%%%%%%%%%%%%%%%%%

\title{Loophole-Free Bell Test Based on Local Precertification of Photon's Presence}
%Jabaquara Project, paper 1.

%%%%%%%%%%%%%%%%%%%%%%%%%%%%%%%%%%%%%%%%%%%%%%%%%%%%%%%%%%%%%%%%%%%

\author{Ad\'an Cabello}
 %\email{adan@us.es}
 \affiliation{Departamento de F\'{\i}sica Aplicada II, Universidad de
 Sevilla, E-41012 Sevilla, Spain}
 \affiliation{Department of Physics, Stockholm University, S-10691
 Stockholm, Sweden}

\author{Fabio Sciarrino}
 %\email{fabio.sciarrino@uniroma1.it}
 %\homepage{http://quantumoptics.phys.uniroma1.it}
 \affiliation{Dipartimento di Fisica, Sapienza
 Universit\`{a} di Roma, I-00185 Roma, Italy}

%%%%%%%%%%%%%%%%%%%%%%%%%%%%%%%%%%%%%%%%%%%%%%%%%%%%%%%%%%%%%%%%%%%

\date{\today}

%First version: August 2011 (Paraty)
%This version: May 14, 2012 (Sevilla), after PRX first proofs

%%%%%%%%%%%%%%%%%%%%%%%%%%%%%%%%%%%%%%%%%%%%%%%%%%%%%%%%%%%%%%%%%%%

\begin{abstract}
A loophole-free violation of Bell inequalities is of fundamental importance for demonstrating quantum nonlocality and long-distance device-independent secure communication. However, transmission losses represent a fundamental limitation for photonic loophole-free Bell tests. A local precertification of the presence of the photons immediately before the local measurements may solve this problem. We show that local precertification is feasible by integrating three current technologies: (i) enhanced single-photon down-conversion to locally create a flag photon, (ii) nanowire-based superconducting single-photon detectors for a fast flag detection, and (iii) superconducting transition-edge sensors to close the detection loophole. We carry out a precise space-time analysis of the proposed scheme, showing its viability and feasibility.
\end{abstract}

%%%%%%%%%%%%%%%%%%%%%%%%%%%%%%%%%%%%%%%%%%%%%%%%%%%%%%%%%%%%%%%%%%%

%Subject Areas: Quantum Physics, Quantum Information, Optics}

%\pacs{03.65.Ud,03.67.Mn,42.50.Xa}
%03.65.Ud Entanglement and quantum nonlocality
%(e.g. EPR paradox, Bell's inequalities, GHZ states, etc.)
%03.67.Mn Entanglement production, characterization, and manipulation
%42.50.Xa Optical tests of quantum theory

\maketitle

%%%%%%%%%%%%%%%%%%%%%%%%%%%%%%%%%%%%%%%%%%%%%%%%%%%%%%%%%%%%%%%%%%%
\section{Introduction}
%%%%%%%%%%%%%%%%%%%%%%%%%%%%%%%%%%%%%%%%%%%%%%%%%%%%%%%%%%%%%%%%%%%

One of the greatest discoveries of modern science is that nonlocal correlations can be established between spacelike separated events \cite{Bell64}. In the last 39 years, many experiments testing Bell inequalities have been performed \cite{FC72,ADR82,KMWZSS95,WJSWZ98,RKVSIMW01,MMMOM08,SUKRMHRFLJZ10}. However, all of them suffer from different limitations, which open loopholes for local-hidden-variable theories. This means that it is possible to construct a local-hidden-variable theory that is able to reproduce the observed data for any of these experiments. This occurs because the experimental conditions do not fully satisfy the hypothesis underlying Bell inequalities. An experimental loophole-free violation of Bell inequalities is of fundamental importance in demonstrating quantum nonlocality, entanglement-assisted reduction of communication complexity \cite{BCMD10}, device-independent secure communication \cite{BHK05,ABGMPS07}, and random-number generation certified by fundamental physical principles \cite{PAMBMMOHLMM10}.

%%%%%%%%%%%%%%%%%%%%%%%%%%%%%%%%%%%%%%%%%%%%%%%%%%%%%%%%%%%%%%%%%%%
% Fig. 1
%%%%%%%%%%%%%%%%%%%%%%%%%%%%%%%%%%%%%%%%%%%%%%%%%%%%%%%%%%%%%%%%%%%

\begin{figure*}[t!!]
\centering
\includegraphics[width=1\textwidth]{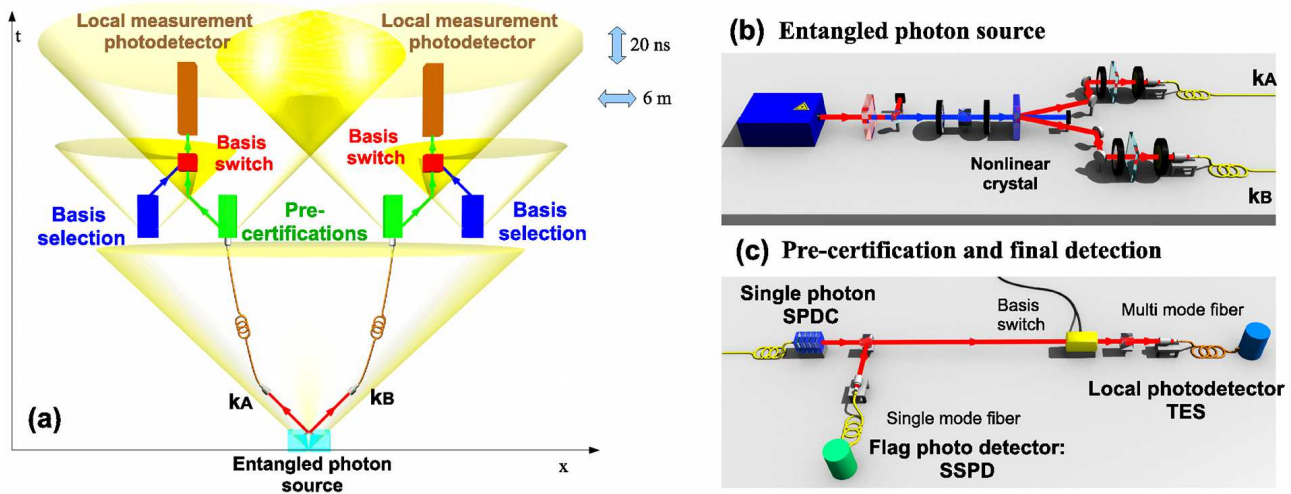}
\caption{\label{Fig1}(a) Space-time diagram of a Bell test based on local precertification of the photon's presence. Green box: precertification stage. Blue box: basis selection. Red box: measurement process. Orange box: final measurement stage. (b) Source of polarization-entangled states. (c) Precertification and detection apparatus.}
\end{figure*}

%%%%%%%%%%%%%%%%%%%%%%%%%%%%%%%%%%%%%%%%%%%%%%%%%%%%%%%%%%%%%%%%%%%

There are two types of loopholes: those related to the separation between the local measurements and those related to the detection efficiency. The locality loophole occurs when the separation between the local measurements is too small to prevent communication at the speed of light between one observer's measurement choice and the other observer's measurement result. To avoid this possibility, these two events must be spacelike separated. The detection loophole occurs when overall detection efficiency $\eta$ is below a certain threshold value. Under this condition, the subensemble of detected events can agree with quantum mechanics, even though the entire ensemble satisfies Bell inequalities \cite{Pearle70}. Therefore, the violation of Bell inequalities only occurs under the assumption that detected events represent the entire ensemble. $\eta$ is the ratio between detected and prepared particles, and is equal to the product of the transmission efficiency, including any losses, and the detector efficiency. For the Clauser-Horne-Shimony-Holt (CHSH) \cite{CHSH69} Bell inequality, the threshold value is $\eta \ge 67\%$ for nonmaximally entangled states \cite{Eberhard93}. For a more complex Bell inequality and ququarts, the threshold is $\eta \ge 62\%$ \cite{VPB10}. Perfect visibility is assumed in both cases; a realistic test would require $\eta \ge 70\%$ \cite{Eberhard93}.

The main problem is that a loophole-free Bell test requires achieving two conditions which are in conflict with each other: long spatial separation and high detection efficiency. In some experiments, the locality loophole has been closed using photons \cite{ADR82,WJSWZ98,SUKRMHRFLJZ10}. In other experiments, the detection loophole has been closed using ions detected with high efficiency \cite{RKVSIMW01,MMMOM08}. However, up until now, no experiment has succeeded in simultaneously close both.

At present, three different approaches are currently being pursued worldwide to achieve a loophole-free Bell test. One is based on the creation of pairs of photons entangled in polarization, free-space distribution along a distance of about 60 m, and detection with high-efficiency superconducting detectors \cite{AJRK06}. However, up until now, this approach has been unsuccessful. Current superconducting transition-edge sensors (TESs) have detection efficiencies beyond $0.97$ \cite{LCPMN10,FFNAYTFIIIZ11}, but to achieve this efficiency the photon has to be inside the fiber ending in the detector in a cryostat operating at low temperature. The propagation of the photon through free space and its coupling with the fiber makes the actual overall photo detection efficiency insufficient for a loophole-free test. This is the so-called transmission problem. The highest reported value in a Bell configuration with (without) spacelike separation is $\eta \sim 38\%$ \cite{WRSLBWUZ11} ($\eta \sim 62\%$ \cite{SGDBFWLCGNW11}). Recently, different schemes have been proposed based on single photon entanglement \cite{SBGRSWH11,CB11,BC12}.

Another optical approach exploits photonic entanglement in continuous variables. The first proposals \cite{NC04,GFCWTG04} used photon substraction on squeezed states to generate quantum states with nonpositive distribution in phase space. However, the violation of the Bell inequality was too small to be observed in practical conditions. Recently, a novel approach combines homodyne measurements (quadrature variables) of high quantum efficiency with photon-number measurements, in order to produce large and, in principle, observable violations of the CHSH inequality \cite{JKLZN10,CS11,CBSSS11}. Very recently, hybrid local measurements combining homodyne measurements and photodetection have been shown to provide violations with arbitrarily low photodetection efficiency, at the cost of a more complex state generation \cite{AQCFCT11}.

A third approach is based on the adoption of entangled states of two remote atoms \cite{RWVHKCZH09}. This requires fast atom detection, which has been recently demonstrated \cite{HKHRWW10}. However, the required spacelike separation between the two entangled systems still represents an experimental challenge. A further possible platform exploits entangled states between atomic and photonic systems \cite{MMBM04,VWSRVSKW06}. Atom-photon Bell experiments in which the atom is detected with certainty only require $\eta \ge 50\%$ using the CHSH inequality \cite{CL07} and $\eta \ge 43\%$ using a 3-setting Bell inequality \cite{BGSS07}.

In this paper, we propose an experimental approach to achieve a loophole-free Bell test. This approach adopts a precertification technique that allows the ``effective'' transmission efficiency to be boosted up to one. Three current technologies will be exploited: (i) enhanced single-photon spontaneous parametric down-conversion (SPDC), (ii) fast nanowire-based superconducting single-photon detectors (SSPDs), and (iii) highly efficient superconducting TESs.

%%%%%%%%%%%%%%%%%%%%%%%%%%%%%%%%%%%%%%%%%%%%%%%%%%%%%%%%%%%%%%%%%%%
\section{Local precertification of photon's presence}
%%%%%%%%%%%%%%%%%%%%%%%%%%%%%%%%%%%%%%%%%%%%%%%%%%%%%%%%%%%%%%%%%%%

The underlying idea is to certify the presence of a photon~$A$ in Alice's location by splitting it into two photons $1$ and $2$ by adopting an enhanced single-photon down-conversion (i). Photon~$2$ is hence detected by a fast detector (ii), and adopted as a flag to precertify the presence of photon~$1$. The configuration is such that photon~$1$ bears the same information initially encoded in photon~$A$. Finally, photon~$1$ is revealed with a highly efficient detector (iii). The advantage of the present scheme is that only the events in which photon~$2$ is detected are involved in the Bell test. In this way, the transmission losses between the entangled source and Alice do not affect the detection efficiency. Moreover, a proper spatial and spectral selection on photon~$2$ can be carried out in order to prepare photon~$1$ in a mode highly coupled with the final detector (iii) at the cost of a reduced experiment rate, but without affecting the final detection efficiency. For two-photon Bell tests, the same conditions should be satisfied to precertify the presence of photon~$B$ in Bob's location. For atom-photon Bell tests, this procedure shall be applied only to the photonic part.

Attention must be paid to possible loopholes due to the rejection of part of the data. Hence, a careful analysis of the space-time diagram has been carried out in Fig.~\ref{Fig1}. Precertification of photon~$1$ (duplication and detection of flag photon~$2$) in Alice's side should not be influenced by Bob's set of local measurements; thus, it must lie outside the light cone of Bob's decision. In addition, to avoid postselection loopholes \cite{CRVDM09} and guarantee that the flag detection is independent of the local settings, Alice's flag photon should be detected before Alice's local setting is established. Therefore, high-speed random-number generators to set the local measurements should be adopted. Moreover, the basis selection should be outside the light cone of the photon duplication, otherwise precertification could influence the basis choice. Finally, photon~$1$ is detected with a detector with resolution time $t_{\rm TES}$ and the measurement process should finish before any signal coming from Bob's location reaches Alice's location.

Let us now address, in more detail, all the different stages introduced above.

{\em (i) Flag photo detectors.---}Current nanowire-based SSPDs \cite{GOCLSSVDWS01,DMBGLMKSKMGLBLF08} are ideal for flagging purposes: They are extremely fast (jitter well below $100$ ps, recovery time below $10$ ns), they have ultralow dark count rates \cite{DMBGLMKSKMGLBLF08}, and they have moderately low detection efficiency ($\eta_{d-{\rm SSPD}}=0.2$ in the visible range, up to $0.60$ for certain wavelengths). While this is not good enough to use them as local-measurement detectors, it is good enough to guarantee a good fraction of precertified events.

{\em (ii) Local-measurement photo detectors.---}Optical TESs have been manufactured for wavelengths of $702$ nm, $800$ nm, $850$ nm, $1310$ nm, and $1550$ nm with detection efficiencies up to $\eta_{d-{\rm TES}}=0.97$--$0.99$ \cite{LCPMN10,FFNAYTFIIIZ11}. Each detector is built for a particular wavelength using a cavity to increase the probability of absorption in the selected wavelength. The typical jitter detection time is $t_{\rm TES}=100$ ns, but recently values up to $10$ ns have been achieved \cite{Lamas11}, while the recovery time is in the $0.1$--$1$ $\mu$s range.

{\em (iii) Enhanced single-photon down-conversion.---}The photon-splitting scheme introduced before can be implemented by adopting the techniques of nonlinear optics. Hereafter, we propose and analyze a scheme based on the process of SPDC. Let us first consider the single photon on mode $A$ to be in the state $\alpha\ket{H}_A+\beta\ket{V}_A$. The photon is injected into a waveguide nonlinear crystal with high nonlinearity. The nonlinear crystal configuration is such that the Hamiltonian of interaction reads
\begin{equation}
 H=i \hbar \chi\left(a_{A,H}a^+_{1,H}a^+_{2,H}+a_{A,V}a^+_{1,V}a^+_{2,V}\right)+h.c.
\end{equation}
There, $a_{i,j}$ stands for the annihilation operator associated to the mode $i$ with polarization $j$. Such a Hamiltonian describes the process in which a single photon is annihilated over mode $A$ with polarization $j$ ($j=H,V$), leading to the generation of a pair of photons, $1$ and $2$, with the same polarization. Energy conservation reads $\nu_A=\nu_1+\nu_2$.

Let us now consider the evolution induced on the incoming single-photon state by the nonlinear process \cite{DS05}. The corresponding unitary evolution reads $U=\exp(-iHt/\hbar)$, with $t$ interaction time, which is applied to the incoming state $(\alpha\ket{1,0}_A+\beta\ket{0,1}_A)\ket{0,0}_1\ket{0,0}_2$, where
$\ket{m,n}_i$ stands for a state with $m$ ($n$) photons with polarization $H$ ($V$) on mode $i$. After the interaction, the output state is
\begin{equation}
\begin{split}
|\Psi\rangle_{A12}=&\cos(g) \left(\alpha|1,0\rangle_A+\beta|0,1\rangle_A\right)|0,0\rangle_1|0,0\rangle_2 \\
&+ \sin(g) |0,0\rangle_A\left(\alpha|1,0\rangle_1|1,0\rangle_2+\beta|0,1\rangle_1|0,1\rangle_2\right),
\end{split}
\end{equation}
where $g$ is the gain of the nonlinear interaction and reads $g=\chi t$. Later, we will comment on the achievable values of $g$ with the present technology. Let us consider the working condition $g=\pi/2$: The single photon at frequency $A$ is deterministically split into two photons, 1 and 2, which, in general, can be nondegenerate in frequency.

In the following step, photons 1 and 2 are divided by using a dichroic mirror, which transmits the photon with wavelength $\lambda_1$ and reflects the photon with wavelength $\lambda_2$. Photon $2$ is measured in the polarization basis $\{\pi_+,\pi_-\}$ and detected with efficiency $\eta$ using the nanowire-based SSPD. When the outcome $\pi_+$ is found, the state of polarization of photon $1$ is $\alpha\ket{H}_1+\beta\ket{V}_1$. The present process does not realize a quantum cloning of the initial qubit $A$. Indeed, while the photon $1$ bears all the information initially encoded in $A$, the measurement of photon $2$ in the basis $\{\pi_+,\pi_-\}$ does not provide any information on the coefficients $\{ \alpha,\beta \}$. The overall process can be interpreted as a quantum nondemolition measurement of the number of incoming photons. Let us now consider that the SPDC crystal is inserted after a lossy channel with transmittance $\eta_t$. If photon $A$ is lost before the SPDC, then no twin-photon generation can occur, since there is no pumping beam exciting the crystal, and hence no photon is detected on mode 2. For the sake of completeness, we address the scenario in which $g$ is different than $\pi/2$. In this case, photon $A$ is split into particles 1 and 2 with probability $\sin(g)^2$. However, since $A$ and 2 have different frequencies, the events in which SPDC did not occur are easily discarded by performing a spectral selection before the detector $D_2$. In summary, the optical scheme described above allows us to herald the presence of a photon in any polarization state after passing through a lossy channel without disturbing the polarization-encoded information.

Let us now apply the same procedure to a polarization-entangled state of two photons, $A$ and $B$. At the end, photons $B$ and $1$ are found in the heralded entangled state: $2^{-1/2}(\ket{H}_B\ket{V}_1+\ket{V}_B\ket{H}_1)$. Clearly, the same nonlinear process can also be adopted on mode $B$ by splitting the photon in two photons $3$ and $4$. Upon heralding by detection of photon $4$, the final state is found to be $2^{-1/2}(\ket{H}_3\ket{V}_1+\ket{V}_3\ket{H}_1)$. We note that the above procedure can also be applied to nonmaximally entangled states that minimize the threshold detection efficiency \cite{Eberhard93}, since it fully preserves the initial quantum states upon heralding.

Let us now estimate the quantitative characteristics of the proposed scheme. The rate at which the two photons are heralded is $R_{\rm exp}=R\mu_C^2\eta_c^2\eta_t^2\eta_{d-{\rm SSPD}}^2$, where $R$ is the rate of the entangled source, $\mu_C =\sin(g)^2$, $\eta_c$ the coupling of the entangled source with single mode fibers, and $\eta_t$ the transmittance up to the single-photon SPDC crystal. This expression takes into account the probability of creating a down-conversion pair both on Alice and Bob's locations, and the probability of detecting the heralding photons on modes 2 and 4. Now the condition for a loophole-free Bell test becomes $\eta_k\eta_{d-{\rm TES}}>0.76$, with $\eta_k$ the coupling of the certified photon with a multimode fiber connected with the TES. Clearly, an efficiency $\eta_k<1$ on the final stage will decrease the final efficiency to $\eta_k\eta_{\rm TES}$. However, the heralded signal photon can be prepared in high-purity state by properly spatially and spectrally selecting the heralding photon. At the final stage, we adopt a multimode fiber for coupling with the TES in order to relax the constraint on the mode preparation and to increase the detection efficiency as much as possible. Previous research mainly focused on heralded photon sources that rely on photon pairs \cite{USBW04,BDGMPPR10}, where one of the photons is used to herald the existence of the other photon within a well-specified spectral bandwidth and spatial single mode. These requirements to have the heralded photon coupled with a single mode fiber and/or narrow bandwidth led to a $\eta_k$ typically within $0.4$--$0.8$ \cite{BRHS10}. The present scheme does not need to fulfill these constraints since the TES is connected with a multimode fiber; hence, we expect the condition $\eta_k=0.8$--$0.9$ to be achievable. The basic idea is that, unlike traditional Bell-type experiments using TESs, we can support a lower experiment rate (due to the filtering process on the heralding photon) with the benefit of improving the final photon coupling. A nonperfect purity of single-photon heralding would not decrease the polarization correlations in the $H$, $V$ basis. Hence, the heralded state will be an entangled state with colored noise for which, as shown below, the minimum detection efficiency exhibits a high robustness against noise.

To address the feasibility of the scheme, we should now estimate which rate $R_{\rm exp}$ can be actually achieved experimentally. A critical parameter is $\mu_C$, which depends on the achievable gain $g$ for the single-photon SPDC. Recently, H\"ubel \emph{et al.}\ reported the observation of SPDC from a single-photon state \cite{HHFRRJ10} (see also \cite{CGR11}). In their apparatus, they adopted a waveguide periodically-poled potassium titanyl phosphate (PPKTP) nonlinear crystal with a length equal to $L = 30$ mm and observed a nonlinearity $\mu_C \approx 10^{-5}$. The adopted crystal was able to generate only pairs of photons with horizontal polarization. For realizing the Hamiltonian necessary for our purpose, a couple of orthogonal crystals should be adopted analogously to those in \cite{KWWAE99}. Attention should be paid to obtaining complete spectral indistinguishability between the two crystals; a possible scheme could adopt a Sagnac configuration, as reported by Fedrizzi \emph{et al.}\ for the generation of narrowband entangled photons \cite{FHPJZ07}. The second feature to be improved is the nonlinearity of the interactions: $\mu_C \propto L^2 d_{\rm eff}^2$. The adoption of commercially available longer crystals and proper coating should easily provide $\mu_C = 4 \times 10^{-4}$ \cite{HJ11}. To fulfill the condition $\mu_C$ of the order of $10^{-3}$, a resonant cavity for the pumping photon could be adopted to improve the interaction strength. An additional factor of $10$ (or even more) is feasible using already demonstrated cavity-based enhancers \cite{OW00,KWOKMUW10}. Moreover, the gain increases if the bandwidth reaching the cavity is narrow; thus, the photon pairs should be selected in a narrow bandwidth first \cite{SRPAHMHHDE10,JRR10}. Other approaches to be pursued could exploit highly nonlinear fibers \cite{HNBHECCIKDFMM07} with a long interaction length $L$ or adopt materials with higher nonlinear coefficients, like organic crystals \cite{JMG08}.

Another possible realization could adopt materials with $\chi^{(3)}$ nonlinearities. The $\chi^{(2)}$ Hamiltonian described above is obtained by adopting one strong pump field. However, higher nonlinearity is achievable due to the classical pump field. Very recently, Langford \emph{et al.}\ observed a value of $g$ equal to $10^{-4}$ \cite{LRPMMZ11}. Higher coupling, in principle even $\mu_C = 1$, could be achieved by increasing the classical pumping beam power.

Let us now estimate the ultimate feasibility of the proposed scheme. We consider a source of polarization-entangled photons with wavelength of $716$ nm based on a nonlinear crystal pumped by an UV pump beam at $358$ nm. The generation rate is set to the value $R=2\times10^7$ s$^{-1}$; this high value can be adopted since the SSPDs exhibit a very high time resolution. The transmittance up to the precertification stage has a typical value $\eta_c\eta_t=0.3$ that includes spectral, spatial filtering, and losses in the channel. We consider a duplication process with efficiency in the range $\mu_C= 2\times10^{-4}$--$10^{-3}$. On both Alice's and Bob's sides, the rates of photons at wavelengths $\lambda_1=1310$ nm and $\lambda_2=1550$ nm read $10^4$ s$^{-1}$. We consider an overall detection efficiency on the flag photon $\eta_{d-{\rm SSPD}}=0.1$ (including spectral and spatial filtering), while the final detection efficiency is close to $\eta_k\eta_{d-{\rm TES}}=0.8$. Adopting the previous values, the final number of estimated events per second is $0.002$--$0.01$ s$^{-1}$, which corresponds to about $10$--$50$ coincidence counts per hour.

Finally, we briefly analyze additional noise introduced by the precertification stages and how it may affect the feasibility of the scheme. The main source of noise can arise in the single-photon SPDC process. In particular, we could expect a residual distinguishability between the emission of photon pairs with horizontal and vertical polarizations. This noise is a so-called colored noise which transforms a nonmaximally entangled state of the form $C |HV\rangle + S |VH\rangle$ into $C^2 \ket{HV}\bra{HV} + (1-p)^2CS (\ket{HV}\bra{VH}+ \ket{VH}\bra{HV}) + S^2 \ket{VH}\bra{VH}$, where $C=\cos(\theta)$ and $S=\sin(\theta)$, and $p$ is the distinguishability between $H$ and $V$ introduced by the single-photon SPDC. The square factor arises because we have two single-photon SPDC processes. Once the noise parameter $p$ is characterized in the experiment, one should determine the strategy which minimize the threshold detection efficiency by properly choosing $\theta$, the measurement to perform on Alice's and Bob's sides, and the Bell inequality to test \cite{CL07,SVCM11}. The minimum threshold detection efficiency for the CHSH inequality has been previously estimated for entangled states with white noise \cite{Eberhard93,CL07,BGSS07}. Here we report preliminary results for the specific noise in our scheme \cite{CFGLSC12}. The interesting point is that this noise almost does not increase the threshold detection efficiency. For example, for a typical $p$ in the $0.01$--$0.04$ range, the threshold detection efficiency is in the $0.676$--$0.702$ range. The robustness of the threshold detection efficiency for colored noise allows us to relax the requirement on the purity of the heralded single-photon states.

Recently, Gisin \emph{et al.}\ \cite{GPS10} have proposed a heralded qubit amplification to implement device-independent quantum key distribution. They introduced a heralded qubit amplifier based on single-photon sources and linear optics that provides a possible solution to overcome the problem of channel losses in Bell tests. Such a scheme could be somewhat related to our approach. While our approach splits a photon in two and detects one photon to precertify the presence of the other one, the protocol of \cite{GPS10} exploits quantum teleportation to realize a heralded single-photon amplifier \cite{RL09}. Hence, one could apply the previous analysis on spatiotemporal separation to perform a loophole-free test. However, the requirement of high indistinguishability for Bell state measurement and the achievement of a sufficiently large $\eta_k$ makes the approach in \cite{GPS10} difficult to implement for a loophole-free Bell test. In \cite{SSCGTZ11}, the possibility of achieving entanglement swapping process to overcome the problem of channel losses by adopting sum-frequency generation has been investigated, while \cite{CM11} considered entanglement swapping with linear optics.

The previous precertification scheme considered photons transmitted over the fiber at $800$ nm. The wavelength of the transmitted pair of entangled photons could be shifted to the telecom ones by adopting single-photon up-conversion at Alice's and Bob's sides \cite{GMD03}. Single-photon up-conversion can have efficiency close to $1$, since the nonlinear interaction exploits a strong pumping beam \cite{TTHABGZ05} and can preserve the information encoded in the incoming single-photon states \cite{GMD03}.

%%%%%%%%%%%%%%%%%%%%%%%%%%%%%%%%%%%%%%%%%%%%%%%%%%%%%%%%%%%%%%%%%%%
\section{Conclusions}
%%%%%%%%%%%%%%%%%%%%%%%%%%%%%%%%%%%%%%%%%%%%%%%%%%%%%%%%%%%%%%%%%%%

We have shown that local precertification of the photon's presence is feasible by integrating three current technologies. The scheme proposed addresses the transmission problem, which is the main obstacle for photonic and atom-photon loophole-free Bell tests, and opens new perspectives for applications. The developments needed to achieve local precertification will also have applications in long-distance photonic communication beyond Bell tests.

%%%%%%%%%%%%%%%%%%%%%%%%%%%%%%%%%%%%%%%%%%%%%%%%%%%%%%%%%%%%%%%%%%%

\begin{acknowledgments}
The authors thank M.\ Bourennane, H.\ H\"ubel, T.\ Jennenwein, A.\ Lamas-Linares, G.\ Lima, P.\ Mataloni, and C.\ Schmiegelow for valuable discussions, and N.\ Spagnolo for discussions and graphical support. This work was supported by the Projects No.~FIS2008-05596 and No.~FIS2011-29400, the Wenner-Gren Foundation, FIRB Futuro in Ricerca-HYTEQ, and Project PHORBITECH of the Future and Emerging Technologies (FET) program (Grant No.~255914).
\end{acknowledgments}

%%%%%%%%%%%%%%%%%%%%%%%%%%%%%%%%%%%%%%%%%%%%%%%%%%%%%%%%%%%%%%%%%%%

%%%%%%%%%%%%%%%%%%%%%%%%%%%%%%%%%%%%%%%%%%%%%%%%%%%%%%%%%%%%%%%%%%%


\begin{thebibliography}{99}

%%%%%%%%%%%%%%%%%%%%%%%%%%%%%%%%%%%%%%%%%%%%%%%%%%%%%%%%%%%%%%%%%%%
% Bell's original Bell paper
%%%%%%%%%%%%%%%%%%%%%%%%%%%%%%%%%%%%%%%%%%%%%%%%%%%%%%%%%%%%%%%%%%%

\bibitem{Bell64}
 J. S. Bell,
 \emph{On the Einstein-Podolsky-Rosen Paradox},
 Physics \textbf{1}, 195 (1964).

%%%%%%%%%%%%%%%%%%%%%%%%%%%%%%%%%%%%%%%%%%%%%%%%%%%%%%%%%%%%%%%%%%%
% Some Bell experiments
%%%%%%%%%%%%%%%%%%%%%%%%%%%%%%%%%%%%%%%%%%%%%%%%%%%%%%%%%%%%%%%%%%%

\bibitem{FC72}
 S. J. Freedman and J. F. Clauser,
 \emph{Experimental Test of Local Hidden-Variable Theories},
 Phys. Rev. Lett. \textbf{28}, 938 (1972).

\bibitem{ADR82}
 A. Aspect, J. Dalibard, and G. Roger,
 \emph{Experimental Test of Bell's Inequalities Using Time-Varying Analyzers},
 Phys. Rev. Lett. \textbf{49}, 1804 (1982).

\bibitem{KMWZSS95}
 P. G. Kwiat, K. Mattle, H. Weinfurter, A. Zeilinger, A. V. Sergienko, and Y. Shih,
 \emph{New High-Intensity Source of Polarization-Entangled Photon Pairs},
 Phys. Rev. Lett. \textbf{75}, 4337 (1995).

\bibitem{WJSWZ98}
 G. Weihs, T. Jennewein, C. Simon, H. Weinfurter, and A. Zeilinger,
 \emph{Violation of Bell's Inequality under Strict Einstein Locality Conditions},
 Phys. Rev. Lett. \textbf{81}, 5039 (1998).

\bibitem{RKVSIMW01}
 M. A. Rowe, D. Kielpinski, V. Meyer, C. A. Sackett, W. M. Itano, C. Monroe, and D. J. Wineland,
 \emph{Experimental Violation of a Bell's Inequality with Efficient Detection},
 Nature (London) \textbf{409}, 791 (2001).

\bibitem{MMMOM08}
 D. N. Matsukevich, P. Maunz, D. L. Moehring†, S. Olmschenk, and C. Monroe,
 \emph{Bell Inequality Violation with Two Remote Atomic Qubits},
 Phys. Rev. Lett. \textbf{100}, 150404 (2008).

\bibitem{SUKRMHRFLJZ10}
 T. Scheidl, R. Ursin, J. Kofler, S. Ramelow, X.-S. Ma,
 T. Herbst, L. Ratschbacher, A. Fedrizzi, N. K. Langford,
 T. Jennewein, and A. Zeilinger,
 \emph{Violation of Local Realism with Freedom of Choice},
 PNAS \textbf{107}, 19708 (2010).

%%%%%%%%%%%%%%%%%%%%%%%%%%%%%%%%%%%%%%%%%%%%%%%%%%%%%%%%%%%%%%%%%%%
% Applications
%%%%%%%%%%%%%%%%%%%%%%%%%%%%%%%%%%%%%%%%%%%%%%%%%%%%%%%%%%%%%%%%%%%

\bibitem{BCMD10}
 H. Buhrman, R. Cleve, S. Massar, and R. de Wolf,
 \emph{Nonlocality and Communication Complexity},
 Rev. Mod. Phys. \textbf{82}, 665 (2010).

\bibitem{BHK05}
 J. Barrett, L. Hardy, and A. Kent,
 \emph{No Signaling and Quantum Key Distribution},
 Phys. Rev. Lett. \textbf{95}, 010503 (2005).

\bibitem{ABGMPS07}
 A. Ac\'{\i}n, N. Brunner, N. Gisin, S. Massar, S. Pironio, and V. Scarani,
 \emph{Device-Independent Security of Quantum Cryptography against Collective Attacks},
 Phys. Rev. Lett. \textbf{98}, 230501 (2007).

\bibitem{PAMBMMOHLMM10}
 S. Pironio, A. Ac\'{\i}n, S. Massar, A. Boyer de la Giroday, D. N. Matsukevich, P. Maunz, S. Olmschenk, D. Hayes, L. Luo, T. A. Manning, and C. Monroe,
 \emph{Random Numbers Certified by Bell's Theorem},
 Nature (London) \textbf{464}, 1021 (2010).

%%%%%%%%%%%%%%%%%%%%%%%%%%%%%%%%%%%%%%%%%%%%%%%%%%%%%%%%%%%%%%%%%%%
% Loopholes
%%%%%%%%%%%%%%%%%%%%%%%%%%%%%%%%%%%%%%%%%%%%%%%%%%%%%%%%%%%%%%%%%%%

\bibitem{Pearle70}
 P. M. Pearle,
 \emph{Hidden-Variable Example Based upon Data Rejection},
 Phys. Rev. D \textbf{2}, 1418 (1970).

%%%%%%%%%%%%%%%%%%%%%%%%%%%%%%%%%%%%%%%%%%%%%%%%%%%%%%%%%%%%%%%%%%%
% The CHSH Bell inequality
%%%%%%%%%%%%%%%%%%%%%%%%%%%%%%%%%%%%%%%%%%%%%%%%%%%%%%%%%%%%%%%%%%%

\bibitem{CHSH69}
 J. F. Clauser, M. A. Horne, A. Shimony, and R. A. Holt,
 \emph{Proposed Experiment to Test Local Hidden-Variable Theories},
 Phys. Rev. Lett. \textbf{23}, 880 (1969).

%%%%%%%%%%%%%%%%%%%%%%%%%%%%%%%%%%%%%%%%%%%%%%%%%%%%%%%%%%%%%%%%%%%
% Detection efficiencies. Symmetric case
%%%%%%%%%%%%%%%%%%%%%%%%%%%%%%%%%%%%%%%%%%%%%%%%%%%%%%%%%%%%%%%%%%%

\bibitem{Eberhard93}
 P. H. Eberhard,
 \emph{Background Level and Counter Efficiences Required for a Loophole-Free Einstein-Podolsky-Rosen Experiment},
 Phys. Rev. A \textbf{47}, R747 (1993).

%%%%%%%%%%%%%%%%%%%%%%%%%%%%%%%%%%%%%%%%%%%%%%%%%%%%%%%%%%%%%%%%%%%

\bibitem{VPB10}
 T. V\'ertesi, S. Pironio, and N. Brunner,
 \emph{Closing the Detection Loophole in Bell Experiments Using Qudits},
 Phys. Rev. Lett. \textbf{104}, 060401 (2010).

%%%%%%%%%%%%%%%%%%%%%%%%%%%%%%%%%%%%%%%%%%%%%%%%%%%%%%%%%%%%%%%%%%%
% Urbana planned loophole-free experiment
%%%%%%%%%%%%%%%%%%%%%%%%%%%%%%%%%%%%%%%%%%%%%%%%%%%%%%%%%%%%%%%%%%%

\bibitem{AJRK06}
 J. B. Altepeter, E. R. Jeffrey, R. Rangarajan, and P. G. Kwiat,
 %\emph{A Loophole-Free Test of Bell's Inequalities},
 in \emph{The 8th International Conference on Quantum Communication, Measurement and Computing (Tsukuba, Japan, 2006). Book of Abstracts},
 edited by O. Hirota, J. H. Shapiro, P. Grangier, A. S. Holevo, P. Kumar, and M. Sasaki (NICT, Tokyo, 2006), p. 158.

%%%%%%%%%%%%%%%%%%%%%%%%%%%%%%%%%%%%%%%%%%%%%%%%%%%%%%%%%%%%%%%%%%%
% TESs
%%%%%%%%%%%%%%%%%%%%%%%%%%%%%%%%%%%%%%%%%%%%%%%%%%%%%%%%%%%%%%%%%%%

\bibitem{LCPMN10}
 A. E. Lita, B. Calkins, L. A. Pellouchoud, A. J. Miller, and S. Nam,
 \emph{Superconducting Transition-Edge Sensors Optimized for High-Efficiency Photon-Number Resolving Detectors},
 Proc. SPIE \textbf{7681}, 7681D (2010).

\bibitem{FFNAYTFIIIZ11}
 D. Fukuda, G. Fuji, T. Numata, K. Ameniya, A. Yoshizawa, H. Tsuchida, H. Fujino, H. Ishii, T. Itatani, S. Inoue, and T. Zama,
 \emph{Titanium-Based Transition-Edge Photon Number Resolving Detector with 98\% Detection Efficiency with Index-Matched Small-Gap Fiber Coupling},
 Opt. Express \textbf{19}, 870 (2011).

\bibitem{WRSLBWUZ11}
 B. Wittmann, S. Ramelow, F. Steinlechner, N. K. Langford, N. Brunner, H. Wiseman, R. Ursin, and A. Zeilinger,
 \emph{Loophole-Free Quantum Steering},
 \eprint{arXiv:1111.0760}.

\bibitem{SGDBFWLCGNW11}
 D. H. Smith, G. Gillett, M. P. de Almeida, C. Branciard, A. Fedrizzi, T. J. Weinhold,
 A. Lita, B. Calkins, T. Gerrits, H. M. Wiseman, S. W. Nam, and A. G. White,
 \emph{Conclusive Quantum Steering with Superconducting Transition-Edge Sensors},
 Nat. Comm. \textbf{3}, 625 (2012).

%%%%%%%%%%%%%%%%%%%%%%%%%%%%%%%%%%%%%%%%%%%%%%%%%%%%%%%%%%%%%%%%%%%
% Proposed Bell experiments with single photons
%%%%%%%%%%%%%%%%%%%%%%%%%%%%%%%%%%%%%%%%%%%%%%%%%%%%%%%%%%%%%%%%%%%

\bibitem{SBGRSWH11}
 N. Sangouard, J.-D. Bancal, N. Gisin, W. Rosenfeld, P. Sekatski, M. Weber, and H. Weinfurter,
 \emph{Loophole-Free Bell Test with One Atom and Less Than One Photon on Average},
 Phys. Rev. A \textbf{84}, 052122 (2011).

\bibitem{CB11}
 R. Chaves, and J. Bohr Brask,
 \emph{Feasibility of Loophole-Free Nonlocality Tests with a Single Photon},
 Phys. Rev. A \textbf{84}, 062110 (2011).

\bibitem{BC12}
 J. Bohr Brask and R. Chaves,
 \emph{Robust Multipartite Bell Tests with a Single Photon},
 \eprint{arXiv:1202.3049}.

%%%%%%%%%%%%%%%%%%%%%%%%%%%%%%%%%%%%%%%%%%%%%%%%%%%%%%%%%%%%%%%%%%%
% Proposed Bell experiments with continuous variables
%%%%%%%%%%%%%%%%%%%%%%%%%%%%%%%%%%%%%%%%%%%%%%%%%%%%%%%%%%%%%%%%%%%

\bibitem{NC04}
 H. Nha and H. J. Carmichael,
 \emph{Proposed Test of Quantum Nonlocality for Continuous Variables},
 Phys. Rev. Lett. \textbf{93}, 020401 (2004).

\bibitem{GFCWTG04}
 R. Garc\'{\i}a-Patr\'on, J. Fiur\'a\v{s}ek, N. J. Cerf, J. Wenger, R. Tualle-Brouri, and P. Grangier,
 \emph{Proposal for a Loophole-Free Bell Test Using Homodyne Detection},
 Phys. Rev. Lett. \textbf{93}, 130409 (2004).

\bibitem{JKLZN10}
 S.-W. Ji, J. Kim, H.-W. Lee, M. S. Zubairy, and H. Nha,
 \emph{Loophole-Free Bell Test for Continuous Variables Via Wave and Particle Correlations},
 Phys. Rev. Lett. \textbf{105}, 170404 (2010).

\bibitem{CS11}
 D. Cavalcanti and V. Scarani,
 \emph{Comment on ``Loophole-Free Bell Test for Continuous Variables Via Wave and Particle Correlations},
 Phys. Rev. Lett. \textbf{106}, 208901 (2011).

\bibitem{CBSSS11}
 D. Cavalcanti, N. Brunner, P. Skrzypczyk, A. Salles, and V. Scarani,
 \emph{Large Violation of Bell Inequalities Using Both Particle and Wave Measurements},
 Phys. Rev. A \textbf{84}, 022105 (2011).

\bibitem{AQCFCT11}
 M. Ara\'ujo, M. T. Quintino, D. Cavalcanti, M. Fran\c{c}a Santos, A. Cabello, and M. Terra Cunha,
 \emph{Bell Tests with Arbitrarily Low Photodetection Efficiency and Homodyne Measurements},
 \eprint{arXiv:1112.1719}.

%%%%%%%%%%%%%%%%%%%%%%%%%%%%%%%%%%%%%%%%%%%%%%%%%%%%%%%%%%%%%%%%%%%
% Proposed Bell tests with atoms
%%%%%%%%%%%%%%%%%%%%%%%%%%%%%%%%%%%%%%%%%%%%%%%%%%%%%%%%%%%%%%%%%%%

\bibitem{RWVHKCZH09}
 W. Rosenfeld, M. Weber, J. Volz, F. Henkel, M. Krug,
 A. Cabello, M. \.{Z}ukowski, and H. Weinfurter,
 \emph{Towards a Loophole-Free Test of Bell's Inequality with Entangled Pairs of Neutral Atoms},
 Adv. Sci. Lett. \textbf{2}, 469 (2009).

\bibitem{HKHRWW10}
 F. Henkel, M. Krug, J. Hofmann, W. Rosenfeld, M. Weber, and H. Weinfurter,
 \emph{Highly-Efficient State-Selective Sub-Microsecond Photoionization Detection of Single Atoms},
 Phys. Rev. Lett. \textbf{105}, 253001 (2010).

%%%%%%%%%%%%%%%%%%%%%%%%%%%%%%%%%%%%%%%%%%%%%%%%%%%%%%%%%%%%%%%%%%%
% Atom-photon Bell tests
%%%%%%%%%%%%%%%%%%%%%%%%%%%%%%%%%%%%%%%%%%%%%%%%%%%%%%%%%%%%%%%%%%%

\bibitem{MMBM04}
 D. L. Moehring, M. J. Madsen, B. B. Blinov, and C. Monroe,
 \emph{Experimental Bell Inequality Violation with an Atom and a Photon},
 Phys. Rev. Lett. \textbf{93}, 090410 (2004); \textbf{93}, 109903(E) (2004).

\bibitem{VWSRVSKW06}
 J. Volz, M. Weber, D. Schlenk, W. Rosenfeld, J. Vrana,
 K. Saucke, C. Kurtsiefer, and H. Weinfurter,
 \emph{Observation of Entanglement of a Single Photon with a Trapped Atom},
 Phys. Rev. Lett. \textbf{96}, 030404 (2006).

\bibitem{CL07}
 A. Cabello and J.-\AA. Larsson,
 \emph{Minimum Detection Efficiency for a Loophole-Free Atom-Photon Bell Experiment},
 Phys. Rev. Lett. \textbf{98}, 220402 (2007).

\bibitem{BGSS07}
 N. Brunner, N. Gisin, V. Scarani, and C. Simon,
 \emph{Detection Loophole in Asymmetric Bell Experiments},
 Phys. Rev. Lett. \textbf{98}, 220403 (2007).

%%%%%%%%%%%%%%%%%%%%%%%%%%%%%%%%%%%%%%%%%%%%%%%%%%%%%%%%%%%%%%%%%%%
% Postselection loophole
%%%%%%%%%%%%%%%%%%%%%%%%%%%%%%%%%%%%%%%%%%%%%%%%%%%%%%%%%%%%%%%%%%%

\bibitem{CRVDM09}
 A. Cabello, A. Rossi, G. Vallone, F. De Martini, and P. Mataloni,
 \emph{Proposed Bell Experiment with Genuine Energy-Time Entanglement},
 Phys. Rev. Lett. \textbf{102}, 040401 (2009).

%%%%%%%%%%%%%%%%%%%%%%%%%%%%%%%%%%%%%%%%%%%%%%%%%%%%%%%%%%%%%%%%%%%
% Flag detectors
%%%%%%%%%%%%%%%%%%%%%%%%%%%%%%%%%%%%%%%%%%%%%%%%%%%%%%%%%%%%%%%%%%%

\bibitem{GOCLSSVDWS01}
 G. N. Gol'tsman, O. Okunev, G. Chulkova,
 A. Lipatov, A. Semenov, K. Smirnov, B. Voronov,
 A. Dzardanov, C. Williams, and R. Sobolewski,
 \emph{Picosecond Superconducting Single-Photon Optical Detector},
 App. Phys. Lett. \textbf{79}, 705 (2001).

\bibitem{DMBGLMKSKMGLBLF08}
 A. Divochiy, F. Marsili, D. Bitauld, A. Gaggero, R. Leoni,
 F. Mattioli, A. Korneev,
 V. Seleznev, N. Kaurova,
 O. Minaeva, G. Gol'tsman, K. G. Lagoudakis, M. Benkhaoul, F. L\'evy, and A. Fiore,
 \emph{Superconducting Nanowire Photon-Number-Resolving Detector at Telecommunication Wavelengths},
 Nat. Photon. \textbf{2}, 302 (2008).

\bibitem{Lamas11}
 A. Lamas-Linares (private communication).

%%%%%%%%%%%%%%%%%%%%%%%%%%%%%%%%%%%%%%%%%%%%%%%%%%%%%%%%%%%%%%%%%%%
% Single-photon down-conversion and enhancement
%%%%%%%%%%%%%%%%%%%%%%%%%%%%%%%%%%%%%%%%%%%%%%%%%%%%%%%%%%%%%%%%%%%

\bibitem{DS05}
 F. De Martini and F. Sciarrino,
 \emph{Non-Linear Parametric Processes in Quantum Information},
 Prog. Quant. Electron. \textbf{29}, 165 (2005).

\bibitem{USBW04}
 A. B. U'Ren, C. Silberhorn, K. Banaszek, and I. A. Walmsley,
 \emph{Efficient Conditional Preparation of High-Fidelity Single Photon States for Fiber-Optic Quantum Networks},
 Phys. Rev. Lett. \textbf{93}, 093601 (2004).

\bibitem{BDGMPPR10}
 A. L. Migdall, S. V. Polyakov, M. Genovese, F. Piacentini, I. Ruo Berchera, I. P. Degiovanni, and G. Brida
 \emph{Experimental Realization of a Low-Noise Heralded Single Photon Source},
 Opt. Express \textbf{19}, 1484 (2011).

\bibitem{BRHS10}
 A. M. Bra\'nczyk, T. C. Ralph, W. Helwig, and C. Silberhorn,
 \emph{Optimized Generation of Heralded Fock States Using Parametric Down-Conversion},
 New J. Phys. \textbf{12}, 063001 (2010).

\bibitem{HHFRRJ10}
 H. H\"ubel, D. R. Hamel, A. Fedrizzi, S. Ramelow, K. J. Resch, and T. Jennewein,
 \emph{Direct Generation of Photon Triplets Using Cascaded Photon-Pair Sources},
 Nature (London) \textbf{466}, 601 (2010).

\bibitem{CGR11}
 M. Corona, K. Garay-Palmett, and A. B. U'Ren,
 \emph{Experimental Proposal for the Generation of Entangled Photon Triplets by Third-Order Spontaneous Parametric Downconversion in Optical Fibers},
 Opt. Lett. \textbf{36}, 190 (2011).

\bibitem{KWWAE99}
 P. G. Kwiat, E. Waks, A. G. White, I. Appelbaum, and P. H. Eberhard,
 \emph{Ultrabright Source of Polarization-Entangled Photons},
 Phys. Rev. A. \textbf{60}, R773 (1999).

\bibitem{FHPJZ07}
 A. Fedrizzi, T. Herbst, A. Poppe, T. Jennewein, and A. Zeilinger,
 \emph{A Wavelength-Tunable Fiber-Coupled Source of Narrowband Entangled Photons},
 Opt. Express \textbf{15}, 15377 (2007).

\bibitem{HJ11}
 H. H\"ubel and T. Jennewein (private communication).

\bibitem{OW00}
 M. Oberparleiter and H. Weinfurter,
 \emph{Cavity-Enhanced Generation of Polarization-Entangled Photon Pairs},
 Opt. Comm. \textbf{183}, 133 (2000).

\bibitem{KWOKMUW10}
 R. Krischek, W. Wieczorek, A. Ozawa, N. Kiesel, P. Michelberger, T. Udem, and H. Weinfurter,
 \emph{Ultraviolet Enhancement Cavity for Ultrafast Nonlinear Optics and High-Rate Multiphoton Entanglement Experiments},
 Nat. Photon. \textbf{4}, 170 (2010).

\bibitem{SRPAHMHHDE10}
 C. Schuck, F. Rohde, N. Piro, M. Almendros, J. Huwer, M. W. Mitchell, M. Hennrich, A. Haase, F. Dubin, and J. Eschner,
 \emph{Resonant Interaction of a Single Atom with Single Photons from a Down-Conversion Source},
 Phys. Rev. A \textbf{81}, 011802(R) (2010).

\bibitem{JRR10}
 Y. Jer\'onimo-Moreno, S. Rodr\'{\i}guez-Benavides, and A. B. U'Ren,
 \emph{Theory of Cavity-Enhanced Spontaneous Parametric Downconversion},
 Laser Phys. \textbf{20}, 1221 (2010).

\bibitem{HNBHECCIKDFMM07}
 K. P. Huy, A. T. Nguyen, E. Brainis, M. Haelterman, P. Emplit
 C. Corbari, A. Canagasabey, M. Ibsen, and P. G. Kazansky
 O. Deparis, A. A. Fotiadi, P. M\'egret, and S. Massar,
 \emph{Photon Pair Source Based on Parametric Fluorescence in Periodically Poled Twin-Hole Silica Fiber},
 Opt. Express \textbf{15}, 4419 (2007).

\bibitem{JMG08}
 M. Jazbinsek, L. Mutter, and P. Gunter,
 \emph{Photonic Applications with the Organic Nonlinear Optical Crystal DAST},
 IEEE J. Sel. Top. Quant. \textbf{14}, 1298 (2008).

\bibitem{LRPMMZ11}
 N. K. Langford, S. Ramelow, R. Prevedel, W. J. Munro, G. J. Milburn, and A. Zeilinger,
 \emph{Efficient Quantum Computing Using Coherent Photon Conversion},
 Nature (London) \textbf{478}, 360 (2011).

%%%%%%%%%%%%%%%%%%%%%%%%%%%%%%%%%%%%%%%%%%%%%%%%%%%%%%%%%%%%%%%%%%%
% Coloured noise
%%%%%%%%%%%%%%%%%%%%%%%%%%%%%%%%%%%%%%%%%%%%%%%%%%%%%%%%%%%%%%%%%%%

 \bibitem{SVCM11}
 F. Sciarrino, G. Vallone, A. Cabello, and P. Mataloni,
 \emph{Bell Experiments with Random Destination Sources},
 Phys. Rev. A \textbf{83}, 032112 (2011).

\bibitem{CFGLSC12}
 G. Ca\~nas, J. F. Barra, E. S. G\'omez, G. Lima, F. Sciarrino, and A. Cabello
 %Detection efficiency for loophole-free Bell tests using local precertification.
 (in preparation).

%%%%%%%%%%%%%%%%%%%%%%%%%%%%%%%%%%%%%%%%%%%%%%%%%%%%%%%%%%%%%%%%%%%
% Recent alternatives
%%%%%%%%%%%%%%%%%%%%%%%%%%%%%%%%%%%%%%%%%%%%%%%%%%%%%%%%%%%%%%%%%%%

\bibitem{GPS10}
 N. Gisin, S. Pironio, and N. Sangouard,
 \emph{Proposal for Implementing Device-Independent Quantum Key Distribution Based on a Heralded Qubit Amplifier},
 Phys. Rev. Lett. \textbf{105}, 070501 (2010).

\bibitem{RL09}
 T. C. Ralph and A. P. Lund,	
 %\emph{Nondeterministic Noiseless Linear Amplification of Quantum Systems},
 in	\emph{Proc. of 9th Int. Conf. on Quantum Measurement and Computing},
 edited by A. Lvovsky
 (AIP, New York, 2009),
 p.~155.

\bibitem{SSCGTZ11}
 N. Sangouard, B. Sanguinetti, N. Curtz, N. Gisin, R. Thew, and H. Zbinden,
 \emph{Faithful Entanglement Swapping Based on Sum-Frequency Generation},
 Phys. Rev. Lett. \textbf{106}, 120403 (2011).

\bibitem{CM11}
 M. Curty and T. Moroder,
 \emph{Heralded-Qubit Amplifiers for Practical Device-Independent Quantum Key Distribution},
 Phys. Rev. A \textbf{84}, 010304(R) (2011).

\bibitem{GMD03}
 G. Giorgi, P. Mataloni, and F. De Martini,
 \emph{Frequency Hopping in Quantum Interferometry: Efficient Up-Down Conversion for Qubits and Ebits},
 Phys. Rev. Lett. \textbf{90}, 027902 (2003).

\bibitem{TTHABGZ05}
 S. Tanzilli, W. Tittel, M. Halder, O. Alibart, P. Baldi, N. Gisin, and H. Zbinden,
 \emph{A Photonic Quantum Information Interface},
 Nature (London) \textbf{437}, 116 (2005).

\end{thebibliography}
\end{document}